# EVIDENCE OF LOCALIZED GROUNDWATER FLOW DURING THERMAL RESPONSE TEST USING DISTRIBUTED THERMAL SENSING

Antoine VOIRAND, Charles MARAGNA.

[1] BRGM, 3 Avenue Claude Guillemin, 45060 ORLEANS

a.voirand@brgm.fr

## ABSTRACT

Four Borehole Heat Exchangers (BHE) were installed at the BRGM Shallow Geothermal Research Facility (BRGM-SGRF) in Orléans, France. They are sixty meters deep and eight meters apart on a line. The facility is localized in the Parisian sedimentary basin. The boreholes are in the Beauce Limestones geological layers, which are locally eroded and karstified. A special type of grouting materiel was used in order to investigate if the boreholes are actually grouted. These measures have confirmed that the four boreholes are grouted along the whole depth. Distributed Thermal Response Tests (DTRT) using optic cables alongside the heat exchangers were performed on each borehole. The distributed thermal sensing, combined with the thermal response test, allows discriminating the heat transfer between geological layers. While the first two boreholes temperature variations with respect to the depth is almost constant, again indicative of a purely conductive heat transfer, the DTS measurement technique gave evidence of a localized enhanced heat transfer on the last two boreholes, between 25 and 40 meters deep. This enhanced heat transfer is believed to be generated by a fast groundwater flow at these depths. What is remarkable is that such a rapid flow can be observed on roughly 15 meters height, while a borehole 8 meters apart exhibits a purely conductive temperature profile. The BRGM has develop a borehole heat exchanger model called SAMBA (Semi Analytical Model for Borehole heat exchanger). It discretizes the heat transfer according to the depth of the borehole, using the thermal resistance and capacity model of Bauer et al. (2011) to model heat transfer inside the borehole, and a moving infinite cylindrical source (MICS) for heat transfer in the ground. Inversion of the SAMBA model for the boreholes exhibiting groundwater flow has been performed to determine the mean ground thermal conductivity and the groundwater flow between 25 to 40 meters deep.

## 1. INTRODUCTION

A thermal response test (TRT) is a method used to determine underground thermal properties, particularly thermal conductivity, for designing borehole heat exchangers in ground source heat pump systems (Spitler et Gehlin 2015). The test involves injecting a defined heat load into a borehole and measuring the resulting temperature changes of the circulating fluid. TRTs have become increasingly popular since the late 1990s and are now routinely used in many countries for designing larger plants, allowing for more reliable borehole sizing (Zhang *et al.* 2021). However, TRTs can be influenced by various factors, including groundwater flow and technical issues, which may affect the accuracy of results (Gehlin et Hellström 2003; Previati et Crosta 2025). A Distributed Thermal Response Test (DTRT) combines distributed temperature sensing with a conventional thermal response test, allowing for more accurate assessment of heterogeneous subsurface thermal conductivity (Rolando *et al.* 2017; Acuña and Palm 2011; Liu *et al.* 2021).

Exploitation of a thermal response test is an inverse method where the parameters of a model are determined to match the model to the measured data. The classical exploitation model is based on the Infinite Line Source theory (ILS), since the test is longer than a few hours, so the impact of the cylindrical nature of the borehole on the heat transfer is less than the heat transfer in the ground, and not long enough for the edge effects at the surface and bottom of the heat exchanger to be significant. The classical TRT duration is about 72 hours, with the first 10 to 20 hours not considered for the test exploitation since the permanent heat transfer regime is not yet fully established. The ILS-based exploitation method assumes a purely conductive heat transfer in a homogeneous and isotropic medium, indeed neglecting groundwater flow effect on the heat transfer. For highly heterogeneous aquifers this discrepancy between the real heat transfer environment and the model on which is based the determining of the ground thermal characteristics can lead to highly over-estimated ground thermal conductivity, and a very different long-term heat transfer dynamic.

## 2. EXPERIMENTAL LAYOUT

### 2.1 Borefield

Four sixty meters deep double-U tube borehole heat exchangers were installed at the BRGM Shallow Geothermal Research Facility (BRGM-SGRF) in Orléans, France. Figure 1 shows the layout of the borefield. To achieve the same pressure drop for each BHE, while spacing them sufficiently to avoid thermal interactions, they were placed in a star configuration, 8 m from a central collector, with the connections buried one meter deep. The inlet and outlet collectors are themselves connected to the machinery building by two pre-insulated DN64 HDPE pipes approximately 20 m long, also buried one meter deep.

Between 20 to 60 meters deep, the boreholes are installed in the Beauce Limestones aquifer geological layers, which are locally eroded and karstified, leading to heterogeneous hydraulic and thermal conductivity.

Two cements were used: a "normal" grout with a thermal conductivity greater than 2 W/m.K, and an "improved" grout with a thermal conductivity greater than 3 W/m.K, marketed by the German cement manufacturer Schwenk. They were chosen for this project because they are enriched with graphite, which greatly increases their magnetic susceptibility. Measuring this susceptibility makes it possible to ensure the presence of cement during and after cementing using a geophysical sensor such as the CemTrakker from Santherr GeothermieTechnik. Such measures were performed three months after cementing and proved that the four boreholes are indeed well cemented.

### 2.2 Metrology

The fluid temperature is measured at the inlet and outlet of each borehole in the field at the collectors via 4-wire PT100 probes installed in stainless steel immersion sleeves. These probes are used to control the heating power injected in the ground via the equation:

$$P = \dot{m} c_p (T_{in} - T_{out})$$

Where $\dot{m}$ is the mass flux in the BHE, and $c_p$ is the mass heat capacity of the fluid.

To perform distributed temperature measurements along the entire height of the probes, an optical cable was installed along one U-tube of each borehole, ensuring that it could not shift from the tube during cementing (Figure 2). These optical cables are welded in series in the collector chamber, then go in the building via the trench at 1 meter deep. A DTS interrogator with Raman effect allows measurements to be taken over the entire borefield in a single interrogation. Halfway between the collector and the building, at the bottom of the trench, next to the fibers, a PT100 temperature sensor was positioned, also connected to the DTS, the temperature being assumed to be constant along the 20 m of trench at 1 m depth. The DTS can thus calibrate the optical fiber measurements in real time by matching the temperature measured by the fiber in the trench with the measurement from the PT100 probe.

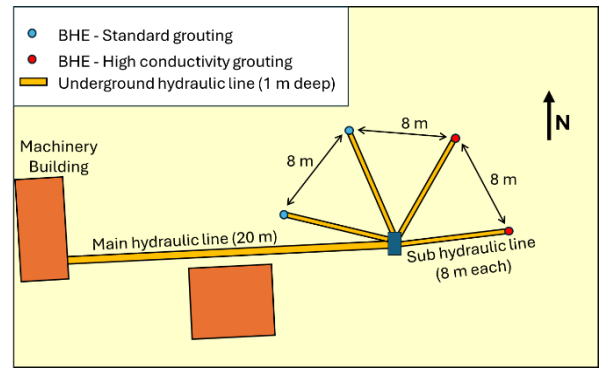

Figure 1: Layout plan of the four borehole heat exchangers.

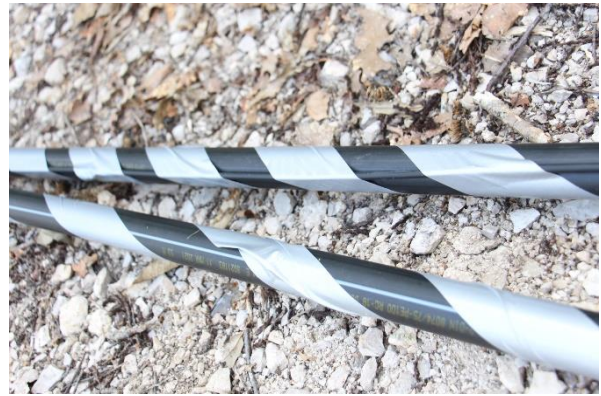

Figure 2: Optical cable fixed on one U-tube of the double-U tube BHE

## 3. THERMAL RESPONSE TESTS

A thermal response test was carried out on each of the BHE installed as part of the Geocoolvert project, the cementing of which was verified and confirmed.

First the fluid is circulated in the BHE without heating for 2 hours, to estimate the undisturbed mean temperature of the ground. Then a constant heat power of 50 W/m of borehole depth (3 kW total heat power) is imposed. Figure 3 shows that the heat power is very stable, except for the transitory period during the first minutes of the tests, and a small undiagnosed variation during the SGV3 TRT which causes a small but visible variation in the mean fluid temperature (Figure 5). The mean deviation from the target total heat power outside of these periods is 6 W (0,2%), with a maximum value of 30 W (1%).

SGV1 and SGV2 BHE exhibit similar purely conductive behavior, with a thermal conductivity of approximately 1,9 W/m.K, which is consistent with the values given by older boreholes close-by on the BRGM geothermal research facility. The higher initial temperature measured on SGV2 is due to an aborted initial TRT on this borehole, carried out 15 days before the start of this new test, and whose residual energy is still present in the ground. This slightly higher initial temperature has apparently no influence on thermal conductivity and borehole resistance measurements. The drilling thermal resistance is also the same for both probes, quite low at approximately 0,06 K.m/W.

These low resistance values are surprising for two reasons. Firstly, because the TRTs were carried out at 1 m$^3$/h with 30% glycol water, which is the working fluid of the thermodynamic machinery of the facility, resulting in a convective resistance between the fluid and the HDPE tube greater than with clear water. Second, because probes SGV1 and SGV2 are cemented with conventional cement at approximately 2 W/m.K, which is not suitable for generating such low thermal resistance.

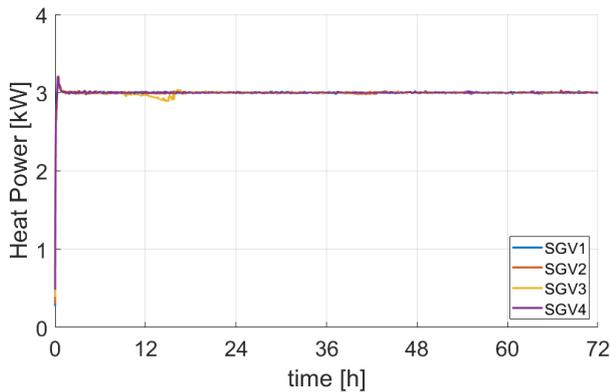

**Figure 3: Heat Power injected in the ground during the TRT of each BHE**

The SGV3 and SGV4 probes, on the other hand, exhibit behavior that is more difficult to interpret in a purely conductive context. Figure 5 shows the mean fluid temperature variation with respect of time on a logarithmic scale, and the values given by the classical purely conductive TRT analysis of these variation. Values of 3,6 and 4,1 W/m.K respectively are not representative of the geological context of the platform. These higher conductivity values are not a priori due to a cementing defect (see section 2.1). They seem rather to indicate the influence of underground flow, bringing a convective term to the heat transfer in the ground, and artificially increasing the apparent thermal conductivity of the ground around the probe.

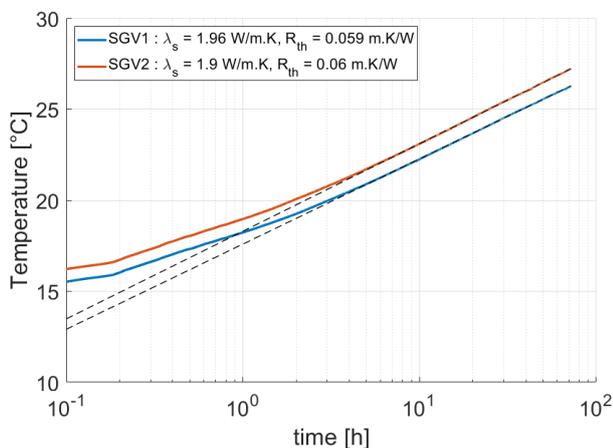

**Figure 4: Mean temperature variation of the fluid during the thermal response test on SGV1 and SGV2, and the result of the classical TRT analysis using ILS model.**

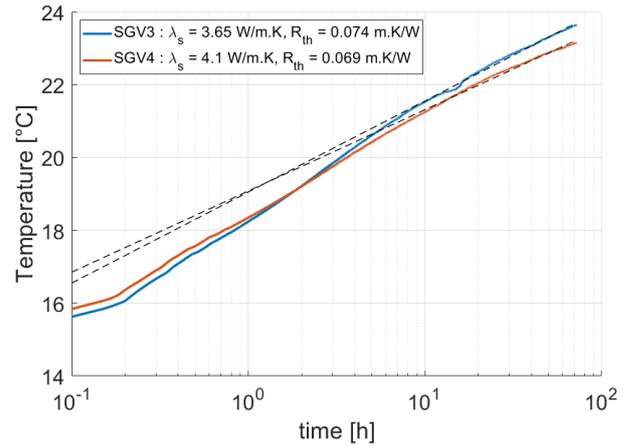

**Figure 5: Mean temperature variation of the fluid during the thermal response test on SGV3 and SGV4, and the result of the classical TRT analysis using ILS model.**

This analysis is supported by the fact that the mean fluid temperature is not a linear function of the logarithm of time after 10 to 20 hours of testing, as opposed to SGV1 and SGV2 results. Excessively high thermal conductivity values lead to excessively high thermal diffusivity values, distorting the calculated borehole thermal resistance values.

Since distributed temperature measurements were taken during the TRT via the optic cables fixed on one U-tube, it was possible to further characterize heat transfer in the four probes.

## 4. DISTRIBUTED THERMAL RESPONSE TESTS

Figure 6 shows the vertical temperature profiles inside boreholes SGV1 to SGV4, at different times during each thermal response test. Because the optical cables are positioned outside the HDPE downcomer and upcomer tubes, the measured temperature is less stable than the heat transfer fluid temperature. It is particularly influenced by the position of the tube relative to the borehole walls and can therefore vary by up to several degrees over a few meters. However, a temperature difference can be observed between the downcomer and upcomer tubes, particularly in the first few meters from the ground.

As expected from the results of conventional TRTs, the temperature profiles inside SGV1 and SGV2 boreholes do not show any significant anomalies. The average tube temperature over depth is relatively constant and increases steadily throughout the test, which is not the case for SGV3 and SGV4. Their profiles indeed exhibit an anomaly between 25 and 40 m depth, particularly pronounced for SGV4. The average fiber temperature in this area varies only slightly between 10 and 71 hours after the start of the test, indicating the presence of significant heat exchange.

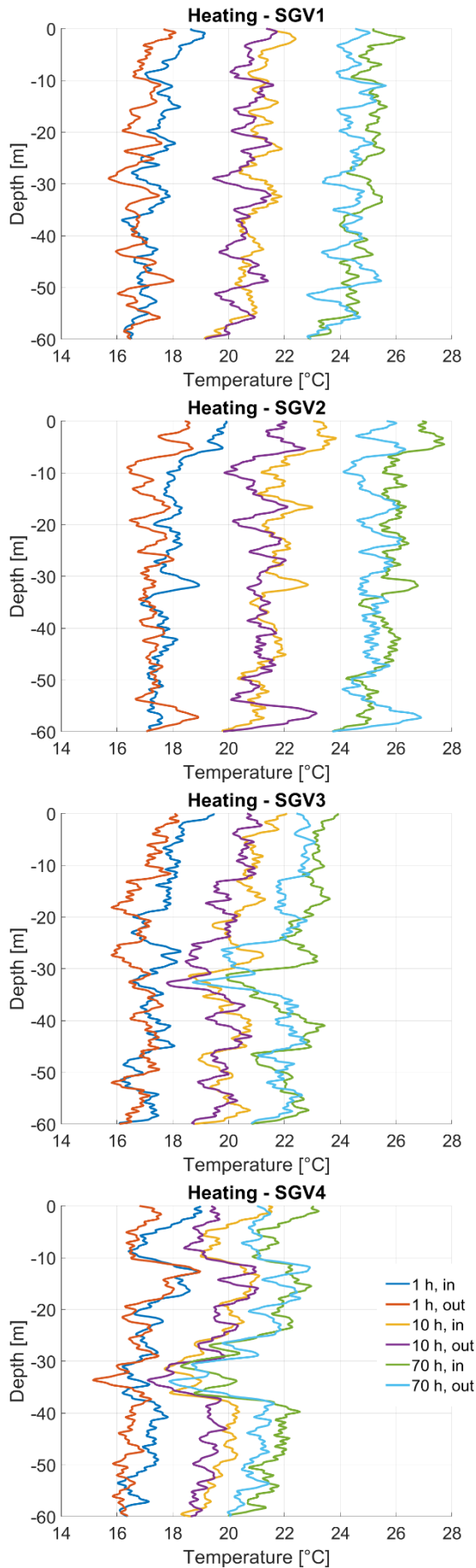

**Figure 6:** Vertical temperature profiles outside the HDPE tube during the heating phase of the TRTs of each BHE. The legend applies to all figures.

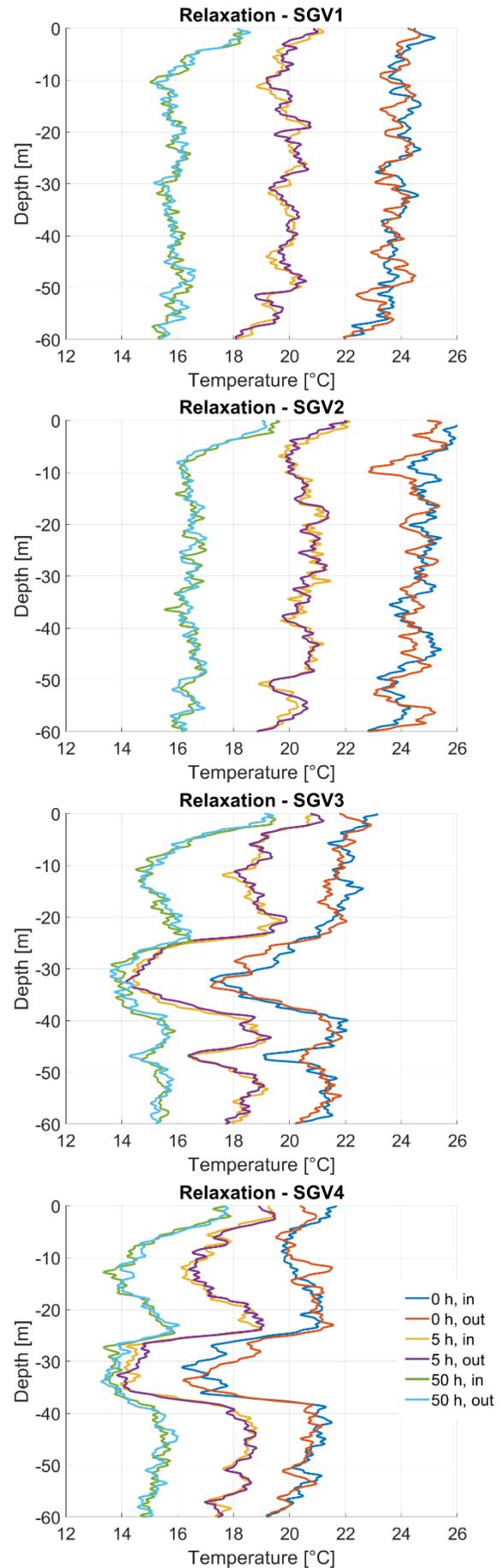

**Figure 7:** Vertical temperature profiles outside the HDPE tube during the relaxation phase of the TRTs of each BHE. The legend applies to all figures.

This result is even more pronounced during the relaxation phase (Figure 7), after the heating phase of the TRT, and circulation of the heat transfer fluid has stopped. The soil is then left without thermal disturbance close to the fiber, which leads to a significant reduction in temperature variations due to the position of the tube in the borehole. It is then possible to see even more precisely the zone of greatest heat exchange on SGV3 and SGV4, between approximately 25 and 40 m. Within a few hours, the temperature inside the BHE has almost returned to its initial level (approximately 14°C), while SGV1 and SGV2 boreholes, as well as the more "classic" conductive zones of SGV3 and SGV4 return more slowly to normal and have a temperature between 15 and 16°C fifty hours after the end of heating. As shown in Figure 1, the SGV3 and SGV4 boreholes, which exhibit such different behaviors over a height of 15 m, are only 8 meters apart from the other boreholes. The cementing monitoring that was carried out on these probes, and which confirmed their good cementing, is of great help in trying to explain this surprising result, since it allows us to exclude the hypothesis of a cementing defect leading to a natural convection phenomenon in the borehole. The explanation is therefore linked to the nature of the surrounding terrain, and not to the installation of the BHE.

It is likely that SGV3 and SGV4 boreholes intersect a large karst conduit (approximately 15 m in diameter). This conduit is not empty, as it could in the Loire region, otherwise the boreholes could never have been cemented properly as shown by the monitoring. However, it is likely to be filled with very eroded and/or cracked rocks, leading to a privileged circulation of groundwater.

## 5. UNDERGROUND HEAT TRANSFER MODEL AND MODEL INVERSION

To determine the velocity of the underground flow in this karst, a BHE model developed at BRGM, called SAMBA, is used. This model allows the discretization of the thermal parameters of the ground and the borehole as a function of depth. This discretization is made possible by calculating the fluid temperature in the probe at each space step, the link between the heat transfer fluid and the temperature at the borehole wall being deduced from an equivalent electrical model taking into account the heat capacity of the grout, developed by (Bauer et al. 2011). The heat transfer in the ground uses a correlation based on a finite element model of an infinite cylindrical source model in translation, also by BRGM during the GECKO project funded by the French research agency ANR (Charles Maragna et Loveridge 2019).

An optimization algorithm on Matlab is used to determine the different parameters of the SAMBA model and minimize the difference of the model with the experimental data from the TRT. SGV1 and SGV2 are modeled with the assumption of an underground water flow of 0 m/s. The mean ground thermal conductivity around boreholes SGV1 and SGV2 and over the entire depth is found to be 1.88 W/m.K and 1.84 W/m.K respectively, compared with the values from the "classical" analysis, of 1.96 W/m.K and 1.9 W/m.K respectively, which is a difference of 2 to 4% (Figures 8 and 9).

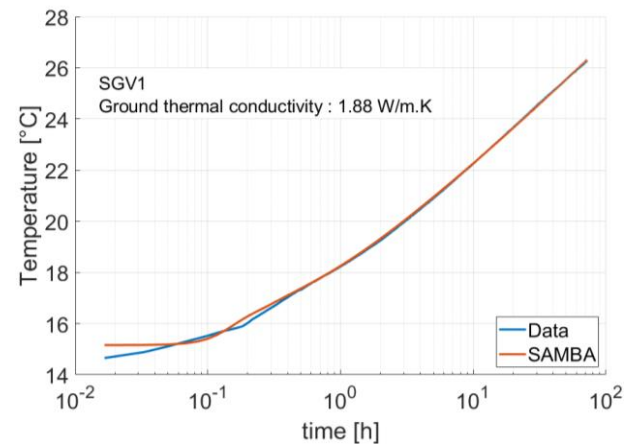

**Figure 8: Comparison between TRT data on SGV1 and SAMBA model with a ground thermal conductivity of 1,88 W/m.K.**

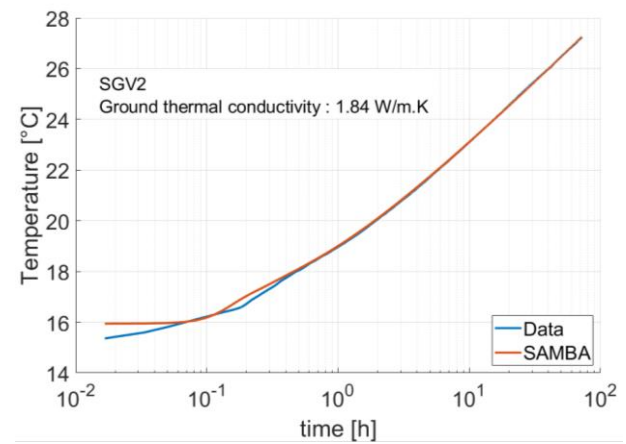

**Figure 9: Comparison between TRT data on SGV2 and SAMBA model with a ground thermal conductivity of 1,84 W/m.K.**

It was possible to work out in the same way, for SGV3 and SGV4, the ground thermal conductivity (averaged over the entire borehole) and the Darcy velocity in the karst conduit between 25 and 40 m depth, which minimizes the difference between the average fluid temperature measured in situ and those calculated by SAMBA (Figures 10 and 11). Although the estimated thermal conductivities are lower than those obtained by a purely conductive interpretation (2.87 and 2.71 W/m.K respectively), they are still high compared to the values obtained for the SGV1 and SGV2 BHE. This is probably due to the fact that the calculated thermal conductivity is averaged over the entire height of the probe. Considering a different conductivity between the conductive zones and the flow zones would probably give more accurate results.

The Darcy velocities allowing to obtain a thermal response identical to that measured are approximately 170 m/year for SGV3, and nearly 300 m/year for SGV4. The hydraulic gradient near the geothermal platform being between 20 and 30 cm per kilometer (Maragna et Maurel 2021), Darcy's law allows to evaluate the hydraulic conductivity of the karst conduit between $2.10^{-2}$ and $5.10^{-2}$ m/s, and therefore a permeability between $2.10^{-9}$ and $5.10^{-9}$ m², while the permeability of the zones outside this conduit is at least an order of magnitude higher.

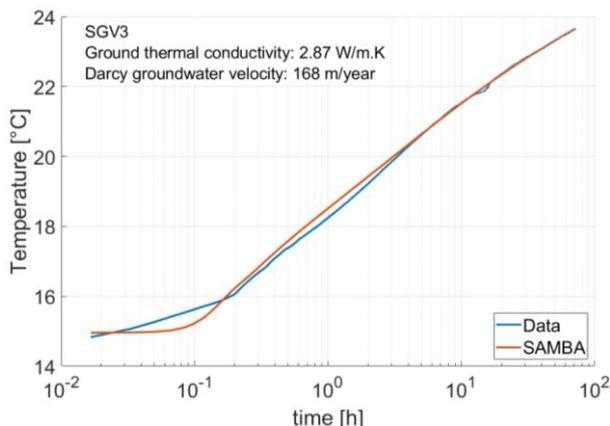

**Figure 10: Comparison between TRT data on SGV3 and SAMBA model with a ground thermal conductivity of 2,87 W/m.K and a Darcy velocity between 25 to 40 m depth of 168 m/year.**

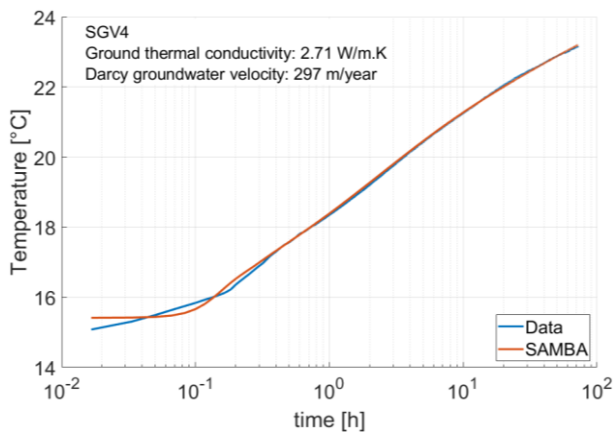

**Figure 11: Comparison between TRT data on SGV4 and SAMBA model with a ground thermal conductivity of 2,71 W/m.K and a Darcy velocity between 25 to 40 m depth of 297 m/year.**

## 3. CONCLUSIONS

Evidence of a localized and important groundwater circulation and its impact on the heat transfer characteristics of a borehole heat exchangers have been observed on four BHE on the BRGM Sallow Geothermal Research Facility in Orléans, France. Two of the four 60 meters depth BHE exhibited conductive heat transfer during thermal response tests, while the other two, placed only 8 m apart, show very strong convective heat transfer between 25 m and 40 m depth. The complete grouting of each borehole has been assessed and confirmed, so the origine of this convective heat transfer should come from an important groundwater velocity around the borehole.

The classical interpretation of the thermal response tests on such ''convective'' BHE does not give a correct estimation of the effective ground thermal conductivity and therefore is not suitable for a proper dimensioning of an energy system using BHE and ground source heat pumps.

BRGM has develop a BHE model taking onto account the parameters variations (ground thermal conductivity, groundwater velocity, tubes positions, etc.) along the depth of the borehole. Interpreting thermal response test of convective BHE is therefore more accurate. However, the number of parameters that can be obtained through the optimization process is limited, which requires assumptions on certain parameters, and suppositions on the spatial variations of certain others.

**Acknowledgements**

This work have been supported by the Centre-Val-de-Loire Region, France, through the research project GEOCOOLVERT.